# Interplay between topological insulators and superconductors

Jian Wang[1,2]*, Cui-Zu Chang[3,4], Handong Li[5,6], Ke He[3], Duming Zhang[1], Meenakshi Singh[1], Xu-Cun Ma[3], Nitin Samarth[1], Maohai Xie[5], Qi-Kun Xue[3,4] and M. H. W. Chan[1]*

[1]The Center for Nanoscale Science and Department of Physics, The Pennsylvania State University, University Park, Pennsylvania 16802-6300, USA
[2]International Center for Quantum Materials, School of Physics, Peking University, Beijing, 100871, China
[3]Institute of Physics, Chinese Academy of Sciences, Beijing 100190, China
[4]Department of Physics, Tsinghua University, Beijing 100084, China
[5]Physics Department, The University of Hong Kong, Pokfulam Road, Hong Kong, China
[6] State Key Laboratory of Electronic Thin Films and Integrated Devices, University of Electronic Science and Technology of China, Chengdu, Sichuan, 610054, China

* Corresponding authors: juw17@psu.edu (Wang); chan@phys.psu.edu (Chan).

Topological insulators are insulating in the bulk but possess metallic surface states protected by time-reversal symmetry. Here, we report a detailed electronic transport study in high quality $Bi_2Se_3$ topological insulator thin films contacted by superconducting (In, Al and W) electrodes. The resistance of the film shows an abrupt and significant upturn when the electrodes become superconducting. In turn, the $Bi_2Se_3$ film strongly weakens the superconductivity of the electrodes, significantly reducing both their transition temperatures and critical fields. A possible interpretation of these results is that the superconducting electrodes are accessing the surface states and the experimental results are the consequence of the interplay between the Cooper pairs of the electrodes and the spin polarized current of the surface states in $Bi_2Se_3$.

**I. INTRODUCTION**

Bismuth-based materials have long been studied for their thermoelectric properties [1-3]. Recently, bismuth selenide ($Bi_2Se_3$), bismuth antimonide ($Bi_{1-x}Sb_x$), bismuth telluride ($Bi_2Te_3$) and antimony telluride ($Sb_2Te_3$) have been predicted theoretically and confirmed experimentally by angle-resolved photoemission spectroscopy (ARPES) experiments to be three dimensional (3D) topological insulators (TIs) due to the strong spin-orbit interactions [1-11]. In transport measurements of 3D TIs [12-19], quantum magneto-resistance (MR) oscillations have been observed and interpreted as evidence of a topologically protected surface state. Of these, $Bi_2Se_3$, with a simple surface state structure (a single Dirac cone) and relatively large band-gap (0.3 eV) has become a reference material in 3D TIs. A key feature of the surface state is that the spin and momentum of the conduction electrons are locked [4], which has been confirmed by ARPES measurements [20, 21] but not directly demonstrated in transport experiments.

In this paper, we report the transport behavior of crystalline $Bi_2Se_3$ films contacted by three different superconducting electrodes to study the interplay between the superconductivity and the TI surface state. We use superconducting bulk indium (In) electrodes and mesocopic aluminum (Al) and tungsten (W) electrodes to study the transport property of the $Bi_2Se_3$ films with thickness of 5 and 200 quintuple layers (QLs) on sapphire and silicon substrates. Every quintuple layer (QL) is one nanometer thick. A simple two-probe configuration is used to minimize the fabrication processing of the electrodes and hence to reduce the risk of altering the intrinsic property of the TI samples. We note that the two-probe (pseudo-four-probe) geometry used here has two contact pads (wires) for each probe. The distances between superconducting electrodes are 1 mm (In) and 1 μm (Al and W). Irrespective of the material of the electrodes, the thickness of the $Bi_2Se_3$ film, the separation of electrodes and the substrates and the contact resistance, the low bias resistance shows a large and abrupt increase near the superconducting transition temperature ($T_C$) of the electrodes. Most interestingly, we observe that the Bi$_2$Se$_3$ films reduce both the $T_C$ and $H_C$ of the superconducting electrodes very significantly.

**II. EXPERIMENTAL RESULTS**

Recent progress in thin film growth of TIs by molecular beam epitaxy (MBE) has made planar TI devices possible [22-26]. Our high quality Bi$_2$Se$_3$ films were grown under Se-rich conditions on sapphire (5 QL) and high resistivity silicon substrates (200 QL) in ultrahigh-vacuum (UHV) MBE systems. A scanning tunneling microscope (STM) image of a 5 QL sample is shown in the left inset of Fig. 1(a). The atomically flat morphology demonstrates the high crystal quality of the film. The carrier density and the mobility of the 5 QL film at 2 K are around $4\times10^{18}$ cm$^{-3}$ ($2\times10^{12}$ cm$^{-2}$) and 3320 cm$^2$/Vs by Hall measurement. With decreasing thickness of the film, the surface to volume ratio increases and surface properties should become more prominent. However, it has been shown by ARPES that in films with thickness less than 5 QL, the interaction between top and bottom surfaces may destroy the topologically protected surface state [23].

The right inset of Fig.1(a) is a schematic diagram of our transport measurement structure. Superconducting In dots of ~ 0.5 mm in diameter and ~ 0.2 mm thick are directly pressed onto the top surface of the Bi$_2$Se$_3$ film. The distance between the two electrodes is ~1 mm. Figure 1(a) shows resistance as a function of temperature (R-T) for the 5 QL Bi$_2$Se$_3$ film. In this paper, unless noted otherwise, the magnetic field is always applied perpendicular to the film and the excitation current for the measurement is 50 nA (corresponding essentially to a zero bias resistance measurement). From 300 K to 45 K, the R-T curve shows linear metallic behavior. A resistance minimum is found near 13.3 K. The residual resistance ratio (RRR) between 300 K and 13.3 K is 2.1. Below 13.3 K, the resistance increases gradually with decreasing temperature. However, at 3.29 K (slightly below the $T_C$ of bulk In (3.4 K)) the resistance shows an abrupt increase. This resistance enhancement is shown in more detail in Fig. 1(b). The resistance at 1.8 K (967.23 Ω) is 2.34 times the resistance when the In electrodes are normal at T=3.4 K. With increasing field, this resistance enhancement decreases rapidly. When the field is 200 Oe the enhancement behavior is totally suppressed. This means the actual crticial field of the In electrodes here is lower than 200 Oe which is the critical field of bulk In at 1.8 K. We interpret the enhancement in R to be a consequence of the onset of superconductivity of the In electrodes; however, it appears the transition temperature and critical field of the In electrodes when contacting the $Bi_2Se_3$ film are slightly below the natural values.

Resistance as a function of the magnetic field (R-H) for the 5 QL $Bi_2Se_3$ film are shown in Fig. 2(a). At 4 K (above $T_C$ of In), the R-H curve shows linear magnetoresistance (MR) from 26 kOe to 80 kOe. Such a linear MR has been attributed to the surface states with the linear energy-momentum correlation [26,27]. However, at 1.8 K (below $T_C$ of In), near zero field the sample exhibits a striking MR peak. Between 0.2 and 9 kOe, well above the critical field of the electrodes, the resistance decreases unexpectedly with field. Upon further increase in the magnetic field, the MR shows the same positive linear behavior as the R-H curve at 4 K.

Figure 2(b) shows MR in small field at different temperatures in more detail. Above $T_C$ of the In electrodes (at 3.4 K and 4.0 K), we observe a positive MR. In addition, there is a small MR dip around zero field, which has been studied carefully and attributed to the weak anti-localization effects in TIs [26,28]. At temperatures below $T_C$ of the In electrodes, the MR dip disappears and a sharp MR peak emerges. With decreasing temperature, the peak value increases rapidly, consistent with the R-T curves of Fig.1. At 1.8 K and 2.4 K, besides the sharp resistance peak around zero field, an additional negative MR is observed from 200 Oe to respectively 9 and 7 kOe. At lower temperature, the negative MR is more robust. This result is unexpected. TI films contacted with normal metal electrodes show positive MR in perpendicular field [28], hence the negative MR cannot be from the TI film itself. On the other hand, if the observed negative MR is due to the superconductivity of indium electrodes, one would not expect this behavior for fields larger than the critical field ($H_C$) of the indium electrodes (~200 Oe at 1.8 K). Interestingly, this negative MR behavior extends up 9 kOe, but only at temperatures below $T_C$ of the electrodes.

Figure 2(c) shows the details of the MR peak shown in Figures 2(a) and (b). Under higher field resolution, the MR 'peak' appears as a plateau with terraces. When we scan magnetic field from negative to positive values and then from positive back to negative values, the sample exhibits hysteretic behavior at 2.6 K and 1.8 K. The plateau/terrace structure of the MR peak and the hysteresis are suggestive of a ferromagnetic response in the conduction electrons. We note that there is no possibility of magnetic contamination in the process of sample preparation. Three dimensional image of the resistance as a function of field and temperature and the resistance contour map along the T-H axes constructed from the experimental data we have obtained on this sample are shown in Fig. 3. More details can be revealed in this figure. In addition to bulk In electrodes measurements, mesoscopic superconducting Al and W electrodes were patterned on the TI films to test the universality of the observed phenomena. The inset of Fig. 4(a) is a scanning electron microscope (SEM) image of our measurement structure. The $Bi_2Se_3$ film for this sample is 200 nm thick and grown on high resistivity silicon substrate. The substrate is completely insulating below 150 K. The carrier density and the mobility of the film at 1.8 K are found to be $2.76 \times 10^{18}$ cm$^{-3}$ ($5.52 \times 10^{13}$ cm$^{-2}$) and 2800 cm$^2$/Vs by Hall measurements. The superconducting Al electrodes are 50 nm thick and directly deposited on the top surface of the film by electron beam lithography (EBL) followed by e-beam assisted evaporation. The distance between the two Al electrodes is 1 μm. Figure 4(a) shows the R-T curves of this sample. Under zero magnetic field, there is a sharp resistance increase at 0.95 K from 23.5 Ω which becomes saturated below 0.85 K at 28.5 Ω. This enhancement is similar to our observation in the sample with bulk In electrodes. The onset temperature of the resistance enhancement at 0.95 K is significantly lower than the $T_C$ of 50 nm thick Al film. This enhancement is suppressed by a field of 100 Oe. An Al film of the same thickness evaporated with the same procedures on an insulating $Si_3N_4$ substrate shows a $T_C$ of 1.4 K and a critical field of ~800 Oe at 0.65 K. Thus, the superconductivity of the Al electrodes is substantially and clearly weakened by the TI film. As shown above, the effect of the $Bi_2Se_3$ film on the bulk In electrodes is not as strong. This is not unreasonable since the Al electrodes are only 50 nm thick and the In electrodes are 'macroscopic' in size. The MR behavior in small field as shown in Fig. 4(b) is similar to that found with In electrodes as shown in Fig. 2(c). The observations in this Al electrode device were confirmed in two additional devices with the same geometry by zero-bias differential resistance measurement, which was carried out using a lock-in amplifier with 100 nA ac excitation at a frequency of 97 Hz.

By means of the focused ion beam (FIB) deposition technique [29-32], superconducting W electrodes was fabricated on the $Bi_2Se_3$ film (inset of Fig. 5(a)). The thickness of the $Bi_2Se_3$ sample is also 200 nm and the distance between two W electrodes is 1 μm. The FIB-deposited amorphous W strips have been used in a number of experiments as superconducting electrodes [29-31]. The $T_C$ of the strips depends on the exact deposition parameters of the FIB process but were found consistently to be between 4 and 5 K when contacting metallic and magnetic nanowires [29-32]. This is much higher than the $T_C$ of pure W (~12 mK) because the FIB-deposited W strips actually contain approximately 40% atomic carbon and 20% atomic gallium [32]. During the FIB deposition process, the top layers of the $Bi_2Se_3$ film are etched away making fresh contact between the electrodes and the film. It has been shown that in the FIB process electrically transparent interfaces were achieved [30]. Figure 5(a) shows the R-T curves of the W-$Bi_2Se_3$-W structure at different fields. The superconductivity-induced resistance enhancement is again seen in this structure. The resistance increases from 0.5 to 6.5 Ω when the W electrodes turned superconducting. The onset temperature of the resistance enhancement is around 3.5 K, which is again significantly smaller than the $T_C$ (4-5 K) of the W strips [29-31]. The magnetic field sufficient to suppress the resistance peak is less than 10 kOe at 2.2 K, which is again much smaller than the $H_C$ of the W strips (~80 kOe) [29,30]. These results confirms the findings with In and Al electrodes that the $Bi_2Se_3$ film weakens the superconductivity of the contacting electrodes. The MR behavior shown in Fig. 5(b) is also consistent with that shown in Fig. 2. For MR scans made below $T_C$ of W, in addition to the prominent peak at low field, a minimum in R is found at a field value above $H_C$ of the specific temperature of the scan.

To further understand the interplay between the topological insulator thin films with superconducting electrodes, we also carried out differential conductance measurements. Figure 6(a) plots field dependent I-V characteristics of the W-$Bi_2Se_3$-W sample at 0.5 K. There is a sudden voltage drop (negative conductance) when the field is less than 1.5 kOe in I-V curves. This phenomenon is not fully understood. Figure 6(b) shows bias dependent differential conductance (dI/dV) of the same sample at different fields at 0.5 K. The negative dI/dV for small field (H < 2 kOe)_is due to the voltage drop in Fig. 6(a). Apart from its negative value, the differential conductance in zero field is also strongly suppressed below 0.33 mV (~Δ/2e of W, where Δ is the energy gap of the FIB deposited W). This differential conductance suppression becomes weaker and moves to lower bias as the field increases. At 30 kOe, the differential conductance switches from a suppression to an enhancement in small bias regime. Finally at 50 kOe, the dI/dV curve becomes a constant, which means the whole system becomes normal. In Fig. 6(c), we map out the variation of dI/dV as a function of current and magnetic field at 0.5 K. In low field and low excitation current regime, the differential conductance is small, which is consistent with the observation of resistance upturn (Fig. 5). At a small current bias, with increasing field, the differential conductance increases firstly, then decreases. The dI/dV measurements further confirms the observed resistance upturn behavior in such a W-$Bi_2Se_3$-W structure (Fig. 5).

**III. DISCUSSION**

The results shown here indicate the phenomena we have observed are universal and reproducible: they are seen with three different kinds of superconducting electrode materials, $Bi_2Se_3$ film thicknesses of 5 and 200 nm and separation of the electrodes of 1 mm as well as 1 μm. The electrodes are attached onto the surface of the $Bi_2Se_3$ film by mechanical pressure (In), by e-beam fabrication (Al) and by the FIB process (W).

If the contact between a superconductor and a normal metal is electrically transparent, the leakage of Cooper pairs into the normal metal can induce superconductivity in the normal metal. Simultaneously, the superconductivity of the superconductor on the other side of the interface can be weakened. This behavior is called the proximity effect [33]. While the observed weakening of the superconductivity of electrodes in our experiment is qualitatively consistent with this effect, the very substantial decrease in $T_C$ and $H_C$ seen in the Al and W electrodes is unprecedented for superconducting electrodes contacting metallic [29], ferromagnetic [30], semiconducting nanowires [34, 35] and 2D graphene system[36]. The observed increase in R of the $Bi_2Se_3$ film when the electrodes turned superconducting is contrary to the conventional proximity effect interpretation and has not been observed in the half-metallic film employing the same measurement configuration [37].

Extensive studies of transport across superconductor/semiconductor interfaces have shown a range of interesting behavior depending upon the transparency of the contact. When the interface is resistive, as typically occurs due to the formation of an interfacial Schottky barrier, an increased zero-bias resistance often accompanies the normal-superconducting transition of the electrode [38]. In past studies of various semiconductor-superconductor interfaces, such changes in resistance were readily understood using the Blonder-Tinkham-Klapwijk (BTK) model and extensions thereof [39], wherein the interface transparency is a key factor in determining the temperature- and bias-dependence of the transport. Our observations however are quite different from those seen in all past measurements of semiconductor-superconductor junctions. First, they are robust against large variations in the transparency of the contacts. We note that the resistances of the $Bi_2Se_3$ film we measured with the In, Al and W electrodes in the normal and the superconducting states are 425, 23.5, 0.5 Ω and 880, 28.5 and 6.5 Ω, respectively. These values indicate that while the contacts with the In electrodes may be slightly resistive, the Al/$Bi_2Se_3$ and particularly the W/$Bi_2Se_3$ interface are electrically transparent. In spite of the differences in the contact resistance of the three different electrodes, the observed phenonema are essentially the same. Second, we find that the upturn in resistance can be strikingly large compared to that seen in past studies of semiconductor-superconductor junctions [38,39] and shows a behavior contrary to expectations from the BTK model. [40] We are particularly surprised by our observation of the huge resistance upturn (1300 percent) for the W electrodes as well as the strong suppression of differential conductance at low bias, thus unexpectedly showing the largest effect for the highest transparency contacts. Since we cannot explain our observations using the BTK model which has been rigorously and extensively applied to a wide variety of semiconductor-superconductor junctions, we propose that the observed phenomenon is connected to the spin-helical surface states of the $Bi_2Se_3$ film and may be a consequence of the entanglement of bulk and surface transport since we cannot exclude the bulk transport channel in our measurements.

## IV. CONCLUSION

To conclude, a possible explanation of our observations reported here is that we are accessing the special property of TI surface state. The spins of the TI surface states are predicted to be helical with fixed spin orientation at a given momentum [41]. In our transport measurement configuration, the collective spin polarization of the TI surface state is aligned by the current [42]. When the electrodes become superconducting, the spin-singlet Cooper pairs are not compatible with the spin-polarized electrons on the TI surface. Spin flip processes must take place at the interface when the Cooper pairs leak from the current source electrode to TI and also when the spin-polarized electrons flow from TI to the superconducting sink electrode. This process produces a sharp resistance enhancement below $T_C$. The spin-polarized current in turn strongly weakens the superconductivity of the superconducting electrodes. Note: Recently transport measurements with superconducting electrodes were also made on $Bi_2Se_3$ nanoribbons [43] and flakes [16, 44], showing proximity effect and a downturn of zero bias resistance. There are two possible explanations for the different behavior between the observatiosn reported here and in nanoribbons/flakes. It is possible that the "minimally processed" samples used in our present study allow a better preservation of the spin-momentum locked surface states. Another possibility is that the measurements of nanoribbons and micron-sized flakes are carried out in a measurement geometry which is clearly different from that used in the present manuscript: the former involves sample edges while the latter does not. Additionally, in the nanoribbon system, the behavior is found to be sensitively dependent on the distance between the two superconducting W electrodes. At 0.94 μm, a conductance peak at zero bias is found, however at 1.55 μm, a resistance enhancement at zero bias is observed [43]. The behavior at the larger separation is not inconsistent with the observation of the resistance upturn in TI films. It is possible that in the film situation, when the distance between two superconducting electrodes becomes shorter (for example ~100 nm), the proximity-induced zero bias conductance peak or resistance drop can be observed too.

Whatever the correct explanation for these observations, we believe that a systematic comparative study between these geometries may hold an important clue about the coupling of superconducting states with TI states, which not only exhibits the nature of the proximity effect at TI and superconductor interface [45], but also offers the platform in searching for Majorana fermions [46, 47].

**ACKNOWLEDGMENTS**

This work was supported by the Penn State MRSEC under NSF grant DMR-0820404, the General Research Fund (No. HKU 7061/10P) and a Collaborative Research Fund (No. HKU 10/CRF/08) from the Research Grant Council of Hong Kong Special Administrative Region, the (Chinese) National Science Foundation and Ministry of Science and Technology of China, the National Natural Science Foundation of China (No. 11174007) and National Basic Research Program (NBRP) of China (No. 2012CB921300). We are grateful to Jainendra Jain, Xiaoliang Qi, Shoucheng Zhang, Liang Fu, Chaoxing Liu, Ying Ran, Mingliang Tian and Lin He for illuminating discussions.

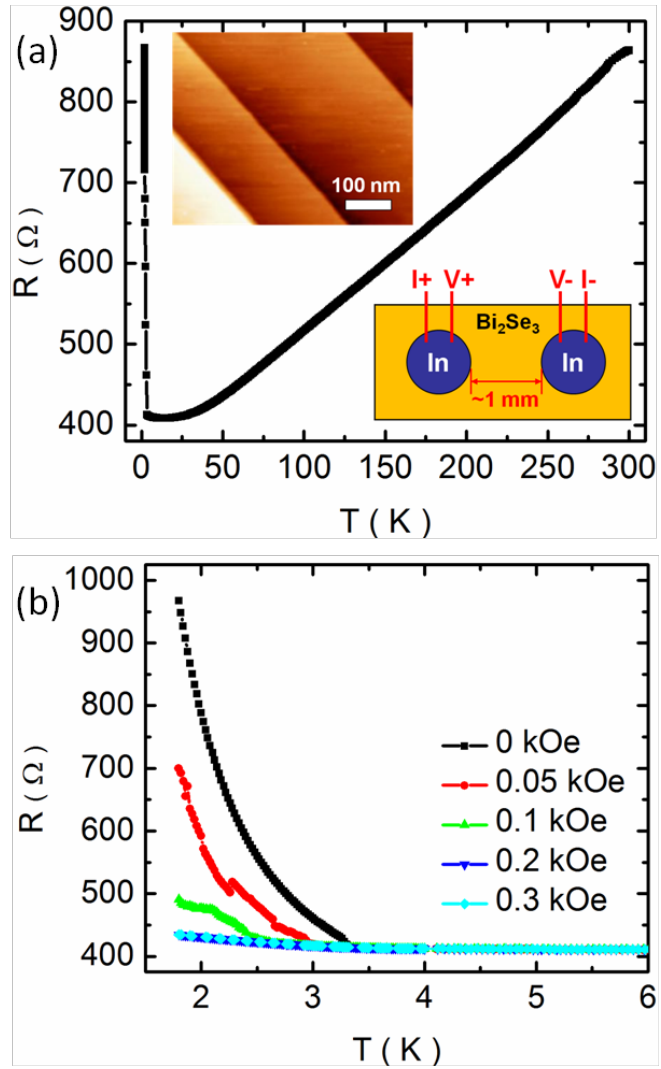

**FIG. 1**. (Color online) R-T behavior of the 5 nm thick $Bi_2Se_3$ film contacted by two superconducting indium dots. (a) Resistance versus temperature of the 5 QL $Bi_2Se_3$ film from room temperature to low temperature. The left inset is a scanning tunneling microscope (STM) image of the $Bi_2Se_3$ film. The right inset is the measurement structure. (b) Resistance versus temperature at different perpendicular fields. The curves at 0.2 kOe and 0.3 kOe are superimposed.

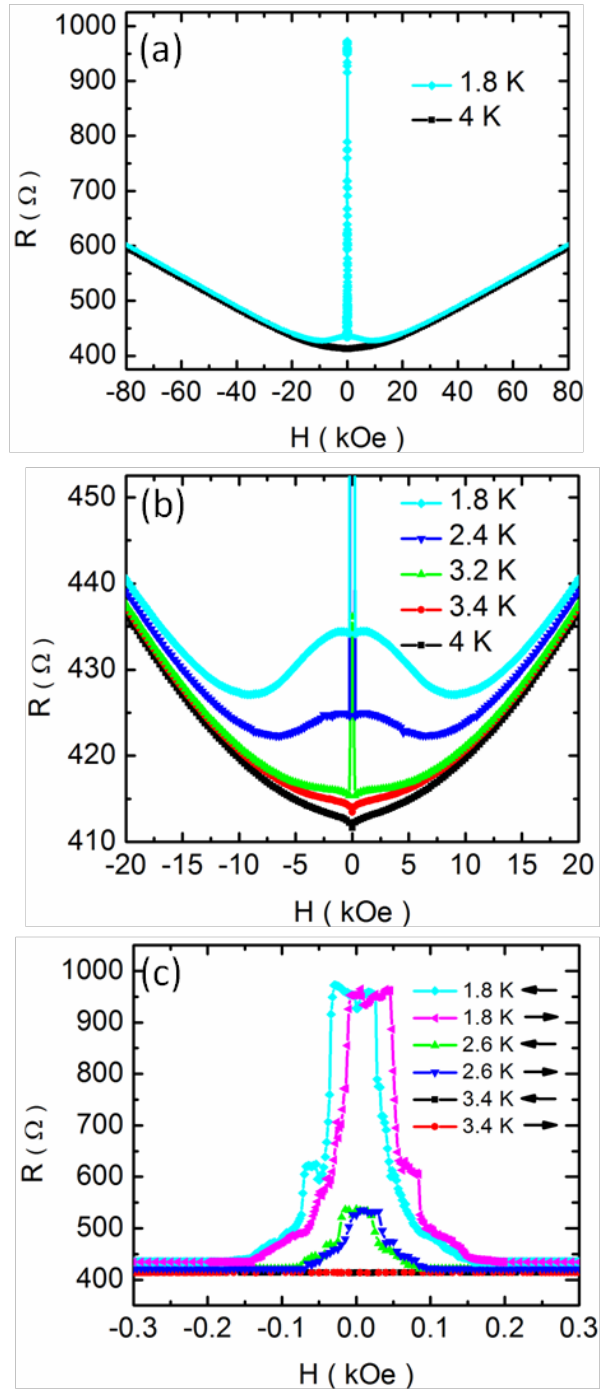

**FIG. 2.** (Color online) R-H scans of the 5 nm thick $Bi_2Se_3$ film contacted by indium electrodes. (a) Resistance as a function of perpendicular magnetic field at 4 K and 1.8 K. (b) Magnified MR for several temperatures. (c) Magneto-resistance peaks near zero field show terrace structure and hysteresis for scans made below $T_C$. Symbol → indicates the scan was made from negative to positive field.

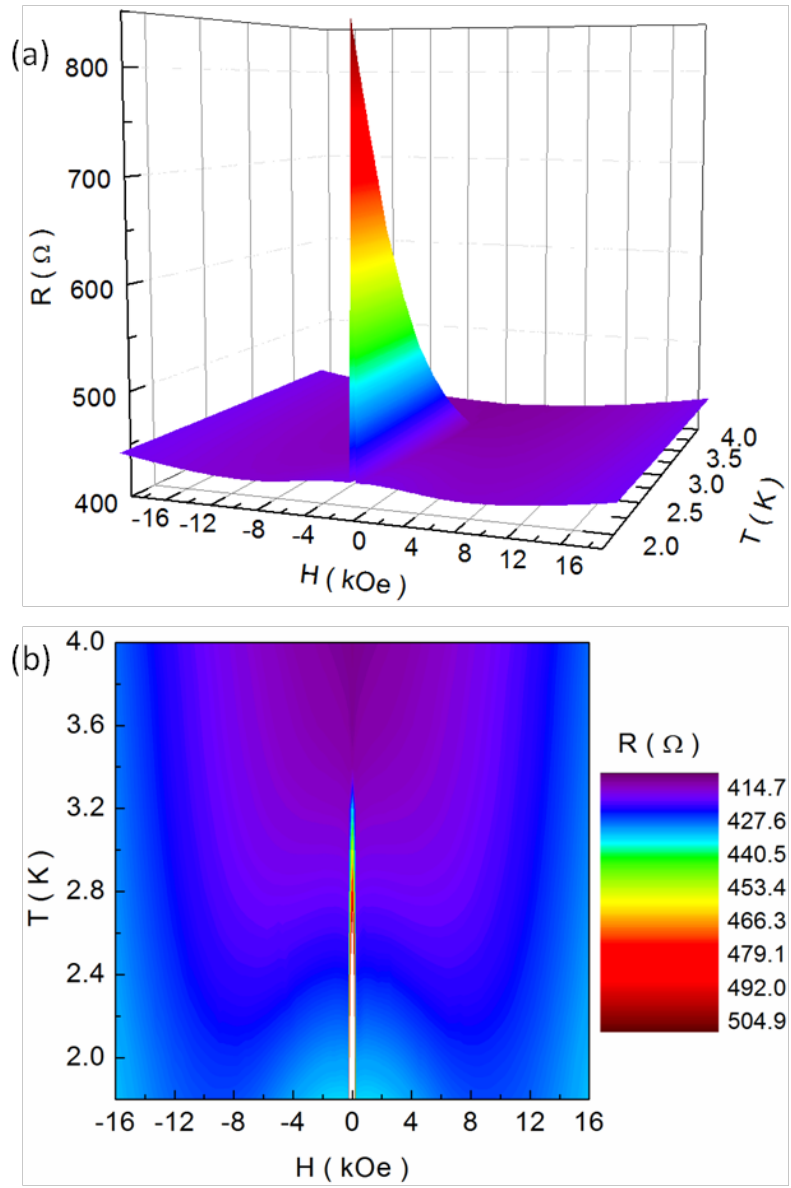

**FIG. 3.** (Color online) (a) Three dimensional image of the resistance as a function of perpendicular field and temperature measured with an bias current of 500 nA for the 5 QL thick $Bi_2Se_3$ film contacted by bulk indium electrodes (In-$Bi_2Se_3$-In). A sharp resistance enhancement induced by the interaction between superconducting electrodes and $Bi_2Se_3$ film is found. (b) Color contour map of resistance along the temperature and perpendicular magnetic field axes of the 5 nm thick $Bi_2Se_3$ film contacted by indium electrodes. The colors represent resistance from 410 Ω (deep purple) to 510 Ω (deep red). The white color means the resistance is larger than 510 Ω.

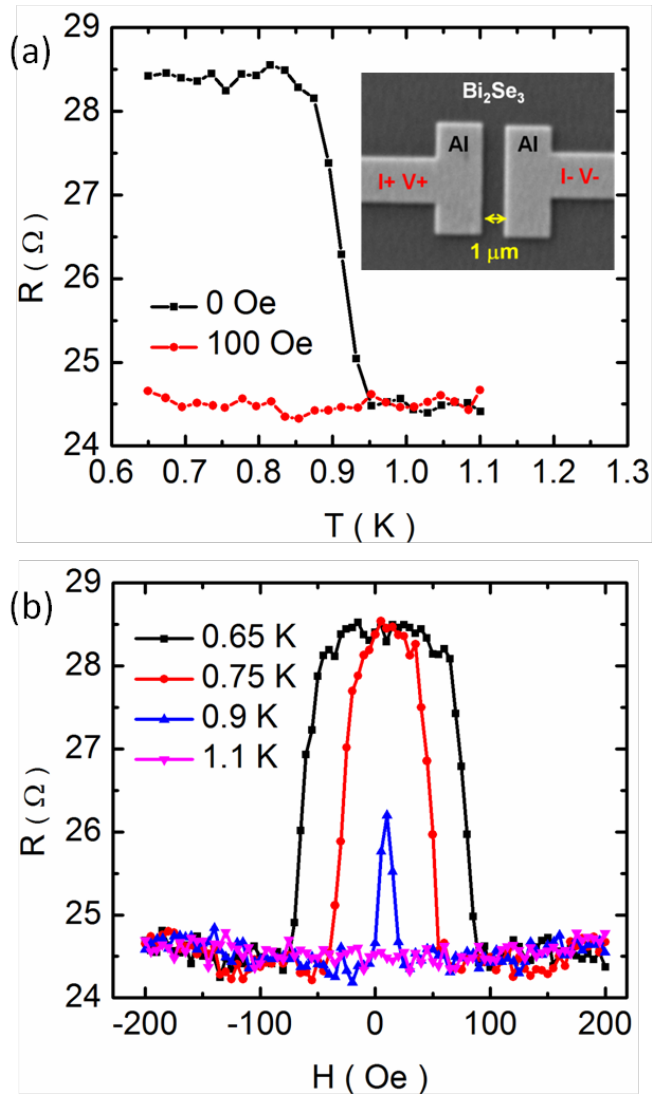

**FIG. 4.** (Color online) Transport behaviors of 200 nm thick $Bi_2Se_3$ films contacted by superconducting Al electrodes. (a) A sharp resistance enhancement is seen at 0.95 K, which saturates below 0.85 K. An applied magnetic field of 100 Oe completely suppresses the enhancement. The inset is a scanning electron microscope (SEM) image of the Al contacts on the surface of the $Bi_2Se_3$ film. (b) The details of the magneto-resistance in small field. The resistance peak at 0.65 K is completely suppressed under a field of less than 100 Oe, which is much smaller than the $H_C$ (800 Oe) of a 50 nm thick Al film not contacting $Bi_2Se_3$ film. $T_C$ of such an 'isolated' Al film is 1.4 K.

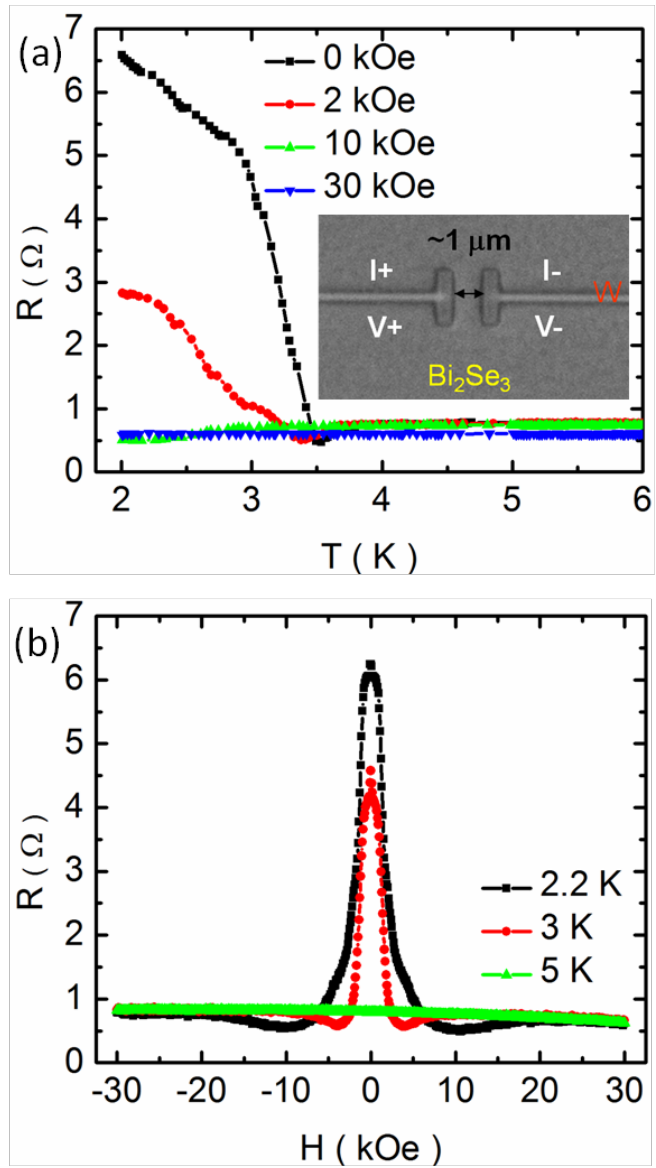

**FIG. 5.** (Color online) Transport behaviors of a 200 nm thick $Bi_2Se_3$ film contacted by superconducting W electrodes. (a) R vs T scans under different magnetic fields. The inset is a SEM image of the W contacts on the surface of the $Bi_2Se_3$ film. (b) Magneto-resistance at different temperatures. When the W electrodes become superconducting, the magneto-resistance shows a large peak around zero field. This behavior disappears when the temperature is larger than $T_C$. The resistance peak is completely suppressed under a field of ~10 kOe at 2.2 K, much smaller than the $H_C$ of the W strips (80 kOe).

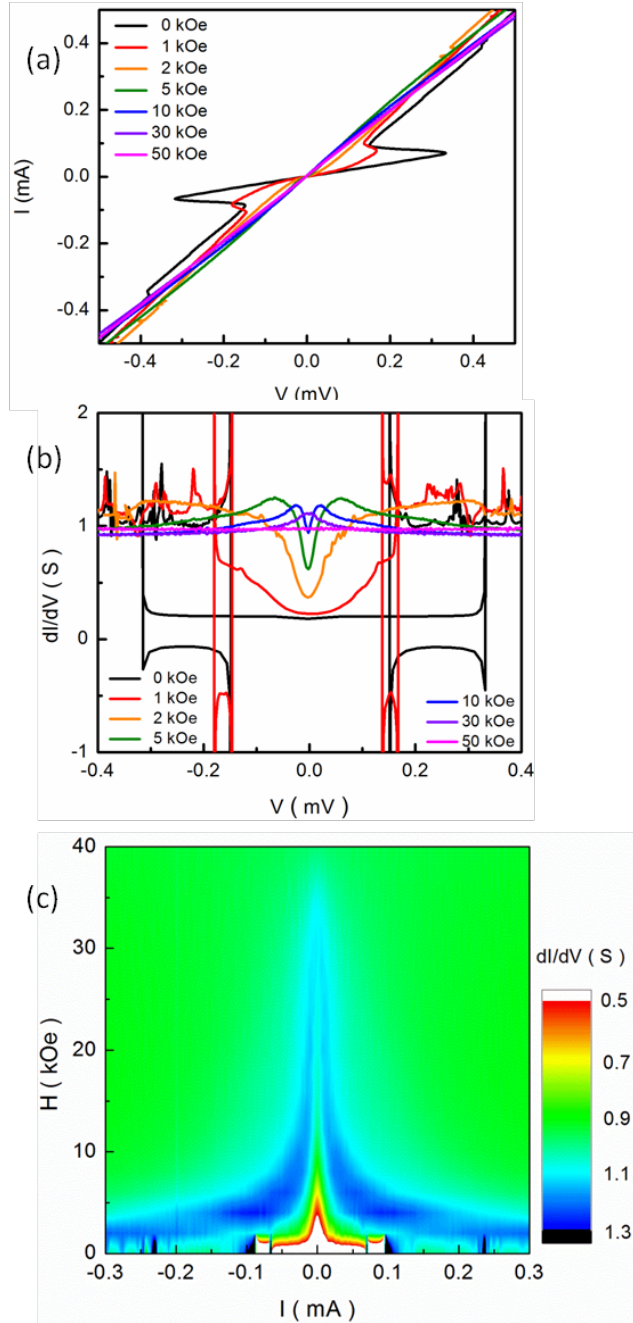

**FIG. 6.** (Color online) (a) Current vs. Voltage at different perpendicular magnetic fields for the W-$Bi_2Se_3$-W sample at 0.5 K. (b) dI/dV vs. V at different magnetic fields at 0.5 K. (c) dI/dV as a function of current I and field H at T = 0.5 K.